\documentstyle[aps,pre,multicol,epsf]{revtex}
\begin{document}

\title{Liesegang patterns: studies on the width law}

\author { M. Droz{}$^1$, J. Magnin{}$^1$, M. Zrinyi{}$^2$ }
\address{ {}$^1$ D\'epartement de Physique Th\'eorique, Universit\'e de Gen\`eve, CH-1211 Gen\`eve 4, Switzerland. }
\address{ {}$^2$ Department of Physical Chemistry, Technical University of Budapest, H-1521 Budapest, Hungary. }


\maketitle

\begin{abstract}
The so-called {\it width law} for Liesegang patterns, which states that the positions $x_n$ and widths $w_n$ of bands verify the relation $x_n \sim w_n^{\alpha}$ for some $\alpha>0$, is investigated both experimentally and theoretically. We provide experimental data exhibiting good evidence for values of $\alpha$ close to 1. The value $\alpha=1$ is supported by  theoretical arguments based on a generic model of reaction-diffusion. 
\end{abstract}
\begin{center}
\bf{UGVA DPT 1999/01-1024}
\end{center}
\begin{multicols}{2}

\section{Introduction}

In recent years, pattern-forming chemical, physical and biological processes came to the foreground. The three main causes of this strong interest are the exotic nature of these processes, designed pattern-forming experiments turned up in physical and chemical laboratories and, last but not least, the dramatic evolution of computing power, allowing to perform extensive numerical simulations as well as to solve numerically the reaction-diffusion differential equations which underly these phenomena~\cite{luthi}.

When an electrolyte diffuses into a gel containing another electrolyte, the eventual formation of a rhythmic pattern of precipitate (a weakly soluble salt) is known as the {\it Liesegang phenomenon}~\cite{liese,Henisch}.
This pattern strongly depends on the dimensionality as well as on the geometry of the experimental setup. 

The simplest and usual way to perform a Liesegang experiment is to take a tube and fill it with a gel which contains the so-called internal electrolyte dissolved in it. The other reactant, refered to as the outer electrolyte, is poured onto the surface of the gel at one extremity of the tube. Usually, the concentration of the latter is much higher than that of the inner one. The main role of the gel in this experiment is to prevent convection of solutions and sedimentation of precipitate. Thus the only mechanism allowed here for the transportation of material is diffusion, i.e. the whole process is diffusion-limited.

Although the Liesegang phenomenon has been known for more than one hundred years, there are still many discussions on the detailed mechanisms responsible for these static structures. Nevertheless, most of the Liesegang patterns one obtains in experiments or observes in nature are characterizable by the following generic laws:

\begin{enumerate}

\item
The ratio between the positions $x_n$, $x_{n+1}$ of two successive bands, measured from the surface of the gel, tends to converge to a finite value that is usually larger than one ({\it spacing law})~\cite{jabli}:

\begin{equation}
\frac{x_{n+1}}{x_n} \stackrel{{\rm n \, large}}{\longrightarrow} 1+p
\end{equation}
with $p>0$.

\item
The ratio between the position of a band and the square root of the time elapsed until it appears, denoted $t_n$, is constant. This law is a direct consequence of the diffusive nature of the dynamics ({\it time law})~\cite{morse}:

\begin{equation}
x_{n} \sim \sqrt{t_n}
\end{equation}

\item It is also generally observed in experiments that the width $w_n$ of the bands increases with their position as a power law ({\it width law})~\cite{widthlaw}:
\begin{equation}
w_n \sim x{_n}{^\alpha}
\label{alpha}
\end{equation}
with $\alpha > 0$.

\end{enumerate}

Until now, a lot of attention has been paid to the spacing law (see \cite{Wagner}, \cite{Prager}, \cite{Zeldovitch}, \cite{usMP}). On the contrary, the width law has largely been ignored for several reasons. Fom the experimental point of view, it is difficult to make precise measurements of the width of the bands without having recourse to sophisticated digitizing methods.
From a theoretical point of view, a complete description of the dynamics of bands formation  requires the knowledge of the detailed mechanisms involved in the coarsening process. Accordingly, two different predictions for $\alpha$ have been previously obtained~\cite{luthi,dee}.

It is our purpose, in this paper, to focus on the width-law aspect of the Liesegang phenomenon, and to try hereby to remove some ambiguities that still surround this basic feature of this class of systems.
In section II, we will expose and discuss the results of Liesegang experiments performed by one of us (M.Z.), for which the widths of the bands were measured. The analysis of the data   provides an accurate estimation of the exponent $\alpha$. In section III, a theoretical argument, independent of the nature of the detailed mechanisms involved in band formation, is developed to support
the experimental findings. Finally, concluding remarks are made in section IV.

\section{Experimental part}

\subsection{Materials and Sample preparation}

We have experimentally studied the formation of  ${\rm Mn(OH)_2}$ precipitate by diffusion of  ${\rm NH_4OH}$  into chemically cross-linked  polyvinyl(alcohol)  (PVA) hydro-gel containing the ${\rm MnSO_4}$ electrolyte. The following reaction occurs when the spatially separated reactants come into contact:
\begin{equation}
Mn^{2+}+2OH^{-} \rightleftharpoons Mn(OH)_2
\end{equation}
\end{multicols}
\begin{figure}[htb]
\centerline{
        \epsfxsize=14cm
        \epsfbox{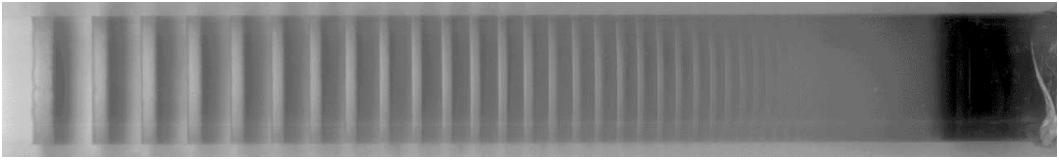}
           }
\vspace{0.5cm}
\caption{Picture of the resulting pattern obtained in one of the three experiments described in Section 2.}
\label{Fig1}
\end{figure}
\begin{multicols}{2}

The highly swollen PVA-hydrogels were prepared by cross-linking of primary PVA -chains with glutaric aldehyde (GDA) in aqueous solution, containing dissolved ${\rm MnSO_4}$. Commercial PVA (Merck 821038) and solution of 25mass\% GDA (Merck) were used for preparation. The initial polymer concentration, as well as the cross-linking density, were kept constant. The polymer content of the PVA gels was in every case 4.0 mass\%. The ratio of monomer unit of PVA (VA) to the cross-linking agent, GDA was: (VA)/(GDA)= 250.
In order to study the precipitation and band formation in swollen networks, one of the reactants, ${\rm MnSO_4}$, was mixed with the polymer solution containing the cross-linking agent. The concentration of ${\rm MnSO_4}$ in the gel phase was  0.03  ${\rm mol/dm^3}$. The gelation was induced by decreasing the pH of the system by ${\rm H_2SO_4}$ (Carlo Erba). The solution was then poured into glass tubes. The gelling process usually took 3-4 hours.
\end{multicols}

\begin{center}
\begin{tabular}{|c|c|c|c|}
\hline
 \hspace{0.5cm} Sample label \hspace{0.5cm} &
 \hspace{0.5cm}$a_0 [mol/dm^3]$ \hspace{0.5cm} &
 \hspace{0.5cm} $b_0 [mol/dm^3]$ \hspace{0.5cm} &
\hspace{0.5cm} $\alpha$ \hspace{0.5cm} \\ \hline \hline
 $\cal A$ & $3.0$ & $0.03$ & 0.94 \\ \hline
 $\cal B$ & $5.0$ & $0.03$ & 0.99 \\ \hline
 $\cal C$ & $14.7$ & $0.03$ & 0.90 \\ \hline
\end{tabular}
\vspace{0.2cm}
\end{center}

Table I.  Concentrations used in each experiment. $a_0$ is the concentration of the the outer electrolyte ($NH_4OH$) and $b_0$, the one if the the inner  electrolyte ($MnSO_4$).  The last column gives the width exponent $\alpha$.
\begin{multicols}{2}
The tubes were sealed, placed vertically,  and allowed to stand undisturbed at a constant temperatures of 17$^{\circ}$C. After completion of the network formation, the gels were brought into contact with the outer reactant, ${\rm NH_4OH}$, thus defining the initial time $t=0$ for the experiment. The concentration of outer electrolyte was varied between 3 and 14.7 $\rm{mol/dm^3}$. We maintained the constancy of boundary condition by refreshing the outer solution continuously. 
Let us underline that due to the choice of the initial conditions, these experiments are effectively one-dimensional. The relevant space coordinate is the one that goes along the longitudinal axis of the tube.

\subsection{Experimental data analysis}

The penetration of  ${\rm NH_4OH}$ into the gel containing ${\rm MnSO_4}$ results in a sharp boundary interface of ${\rm Mn(OH)_2}$ precipitate bands. In many cases we obtained more than 20 bands.
Figure 1 shows a typical example.

In order to determine the space coordinate of the bands obtained, we used a digital video system. A CCD camera with an 1/3" video chip was connected to a PC through a real-time video digitizer card. We also used a scanner to digitize the experimental results.

We first determined the coordinate of the gel surface. Then, the positions of the N bands, $\{ x_n \} _{n=1..N}$, were evaluated. Note that $x_n$ stands here for the position of the upper limit of the precipitate zone. We finally measured the positions $\{ \tilde{x}_n \} _{n=1..N}$ of the lower boundaries of the precipitate layers. The difference of these two quantities provides the width of the n-th band : $w_n \equiv \tilde{x}_n - x_n$. 

It must be mentioned that due to  the sharp boundaries of the precipitate regions, the uncertainties on the width values  are  estimated ranging from 2\% to 5\%.  

\subsection{Experimental results}

Figures 2-4 show the results of three different Liesegang experiments for which
the  concentrations of the electrolytes are given in Table 1. 
They take the form of log-log plots relating the widths $w_n$ to the 
positions $x_n$ of the bands, as measured from the digitized images. The dotted lines are least square linear fits whose slope gives the width exponent $\alpha$. These slope values are shown in Table 1. For all the samples, the width exponent is near $\alpha=1$.

\end{multicols}

\begin{figure}[htb]
\centerline{
        \epsfxsize=7.2cm
        \epsfbox{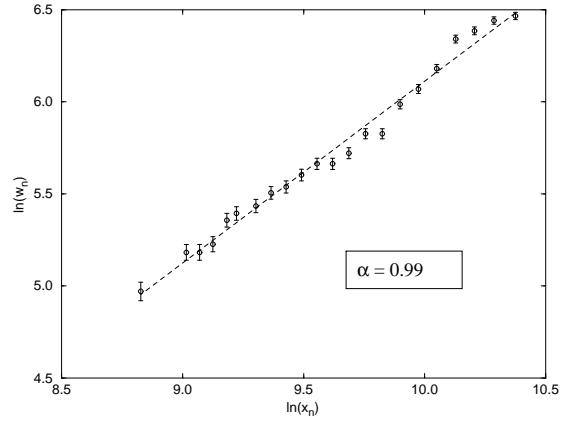}
           }
\vspace{0.5cm}
\caption{Log-log plot of the widths $w_n$ of the bands versus their positions $x_n$ for the sample $\cal A$.  Lengths are in arbitrary units. The dotted line corresponds to a least square linear fit.}
\label{Fig2}
\end{figure}



\begin{figure}[htb]
\centerline{
        \epsfxsize=7.2cm
        \epsfbox{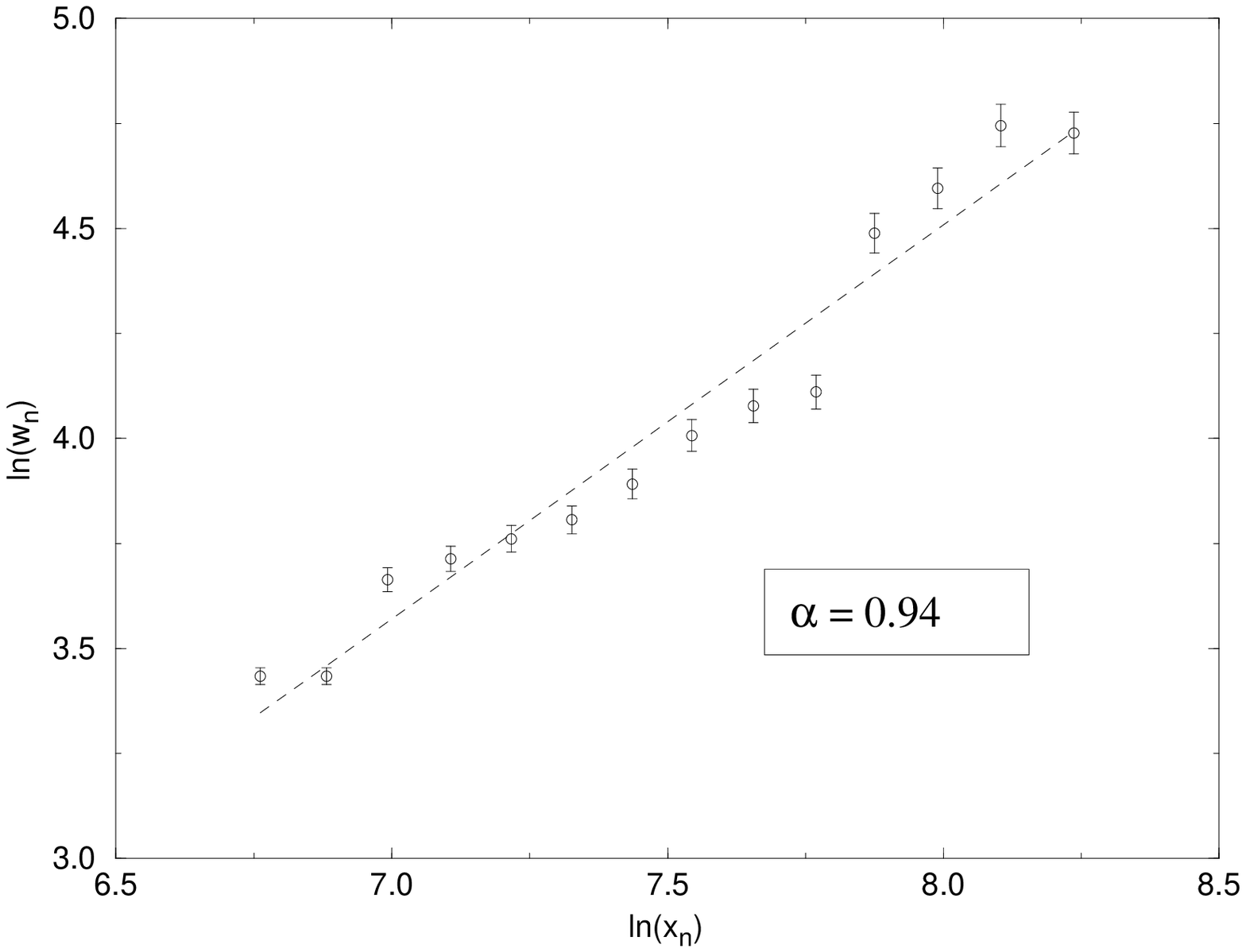}
           }
\vspace{0.5cm}
\caption{Log-log plot of the widths $w_n$ of the bands versus their positions $x_n$ for the sample $\cal B$.  Lengths are in arbitrary units. The dotted line corresponds to a least square linear fit.}
\label{Fig3}
\end{figure}


\begin{figure}[htb]
\centerline{
        \epsfxsize=7.2cm
        \epsfbox{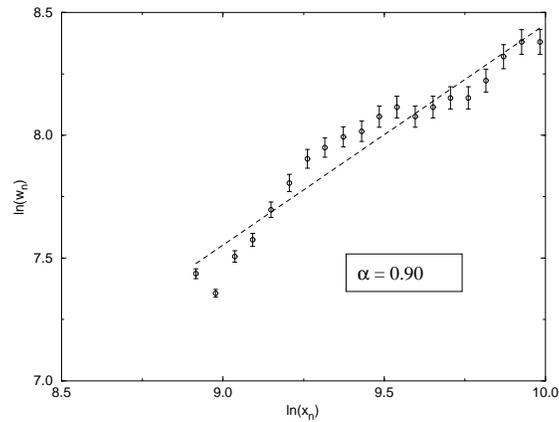}
           }
\vspace{0.5cm}
\caption{Log-log plot of the widths $w_n$ of the bands versus their positions $x_n$ for the sample $\cal C$.  Lengths are in arbitrary units. The dotted line corresponds to a least square linear fit.}
\label{Fig4}
\end{figure}
\begin{multicols}{2}

In these experiments, the two chemical species react and produce an intermediate compound (called  $C$) which also diffuses inside the gel. Its associated traveling front can be put in evidence by light transmission techniques.

\section{Theoretical approaches}

\subsection{Two classes of competing theories for Liesegang patterns}

The several mechanisms one can consider for explaining the formation of 
Liesegang patterns will be, in the context of this paper, 
classified in two categories \cite{usMP}.
Both have in common the following set of ``free'' parameters :
\begin{itemize}
\item{i).} The concentration of the outer reactant A, denoted $a_0$. It is
      generally kept
      fixed at the open extremity of the tube during the whole duration of
      the experiment. It therefore leads to the following initial-and-boundary
      condition for the field $a$ describing the concentration of species A :
      \begin{eqnarray}
      a(0 <x \leq L,t=0) = 0 \\
      a(x=0,t \geq 0) = a_0
      \end{eqnarray}
       where $L$ stands here for the length of the test-tube hosting the gel.

\item{ii).}The initial concentration of the inner electrolyte, $b_0$. One has:
      \begin{eqnarray}
      b(0 \leq x \leq L,t=0) = b_0
      \end{eqnarray}
\item{iii).} The diffusion coefficients of both species in the gel, $D_a$ and $D_b$.
\end{itemize}
      
Each of these two classes of models we distinguish here try  to capture a certain family of microscopical mechanisms leading to the formation of bands. In this respect, they both can be characterized by their own set of parameters associated to these particular mechanisms : 

\begin{enumerate}

\item 
In the first category, we put the {\it ion-product supersaturation}
type theories. They have in common the fact that the outer
and inner reactants directly turn into precipitate where their local 
concentration product reaches some threshold $q^*$. One can symbolize this 
process by the reaction scheme $A+B \rightarrow \underline{D}$ where $\underline{D}$ 
stands here for the precipitate state. The typical and most studied scheme of 
this type is Prager's theory for periodic precipitation \cite{Prager}, 
firstly proposed by Wagner \cite{Wagner}, and investigated later also by  
Zel'dovitch et al \cite{Zeldovitch}. 

\item The second category contains what we will call the {\it intermediate
species-type} theories. We group into this set all 
the schemes that involve the existence of an intermediate state, usually  
denoted by the letter $C$, prior to the formation of the precipitate zones. This compound is supposed to be produced continuously, with a certain rate 
proportional to the local concentration product of the reactants. It thus 
gives rise to a front of $C$ particles that propagates in the system. Such
processes can be tagged with the following symbolic reaction scheme : $A+B 
\rightarrow C \rightarrow \underline{D}$.
It is natural to suppose that this intermediate compound also diffuses inside the gel medium, with a diffusion coefficient $D_c$. \

We recently considered two particular schemes of this type in a 
paper mainly devoted to a deeper understanding of the spacing law 
\cite{usMP}. The first one, called the {\it nucleation and growth} theory, 
supposes that the intermediate compound undertakes a local nucleation 
process at the point where its concentration is greater or equal to a 
certain {\it nucleation threshold} $c^*$. An extension of this theory 
to a case where counter-ions associated to the flow of the outer reactant 
play a catalytic role in the creation of a new band was also examined 
under the denomination of {\it induced sol-coagulation} theory. 

\end{enumerate}

As we noticed in Section II, the experiments we performed show unambiguously the presence of an intermediate $C$ compound. We will therefore base our theoretical considerations on the class of models including the $C$ species. 

\subsection{Prediction of the intermediate-species class for the width law}

The determination of the width law for a particular model with intermediate species can be addressed at different levels. 

A detailed description of the dynamics of band formation can be attempted. However, this route requires a precise knowledge of the microscopical mechanisms as aggregation, flocculation, etc. present in the system, which is clearly too ambitious a goal : very few informations are usually available about which of these mechanisms are present in a given experimental setup.

Nevertheless, if one is only interested in general properties of the bands system, as its width exponent $\alpha$, a qualitative analysis based on general properties like conservation laws should already provide a good approximation for the width exponent. 

We are therefore in search of a general feature of all these intermediate species models that could allow us to compute $\alpha$. Along this line, we found a general argument based on the shape and dynamics of the $C$ front~\cite{usMP}.

Let $t_n$ be the time at which band number $n$ starts to appear. Over the interval $[t_n,t_{n+1}]$, the $C$ profile adopts a typical shape characterized by a diffusive tail at the right of its (growing) maximum, and a quasi-linearly decreasing shape at the left, as shown in Figure 5. This fact can be easily verified by solving numerically the system of partial differential equations modeling the simplest intermediate-species theory producing Liesegang bands, namely :
\begin{eqnarray}
\partial_t a(x,t) & = & D_a \partial_{x}^{2} a(x,t) - ka(x,t)b(x,t) 
\label{modelCbegin} \\
\partial_t b(x,t) & = & D_b \partial_{x}^{2} b(x,t) - ka(x,t)b(x,t) \\
\partial_t c(x,t) & = & D_c \partial_{x}^{2} c(x,t) + ka(x,t)b(x,t) - \\
                  &   & \kappa_1 \Theta(c-c^*)c(x,t) - \kappa_2 c(x,t)d(x,t) \nonumber \\
\partial_t d(x,t) & = & \kappa_1 \Theta(c-c^*)c(x,t) + \kappa_2 c(x,t)d(x,t)
\label{modelCend}
\end{eqnarray}
where $k$,$\kappa_1$ and $\kappa_2$ are reaction rates ($\kappa_1, \kappa_2 \gg 1$), and $\Theta$ is the Heaviside step function. 

Note that in this simple basic model, we just require that $C$ is rapidly going to zero wherever some precipitate already exists. This ingredient is aimed here at including crudely all the microscopic effects like aggregation. 

To discuss the width-law problem, we will balance the quantity of $C$ diffusing out of the $C$ front with the total amount of matter inside a band after its completion.

Let's begin by summing up all the contributions to the quantity of C entering band $n$, once it is formed at $x=x_n$. They can be divided in two parts :

\end{multicols}
\begin{figure}[htb]
\centerline{
        \epsfxsize=8cm
        \epsfbox{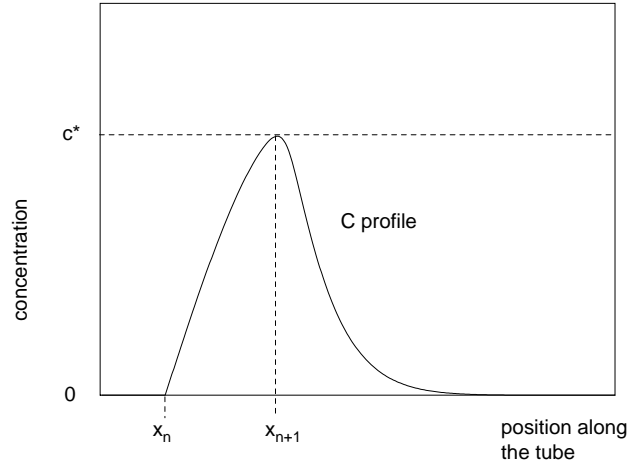}
           }
\vspace{0.5cm}
\caption{Shape of the C profile just before band number $n+1$ appears. $x$ is the coordinate along the tube.}
\label{Fig5}
\end{figure}
\begin{multicols}{2}

Firstly, we can evaluate the quantity of $C$ that flows into band number $n$ during the time interval $(t_n,t_{n+1})$. This quantity, denoted here $Q_{n,1}$, is equal to the time integral over the flux of $C$ toward $x_n$, evaluated from $t_n$ to $t_{n+1}$. This flux is proportional to the slope of the $C$ profile at the right of $x_n$. 
Numerical estimations of this slope show that it scales with time as ~\cite{usMP}:
\begin{equation}
\partial_x C(x=x{_n}{^+},t_n<t<t_{n+1}) \sim \frac{1}{\sqrt{t-t_n}}
\end{equation}
We therefore have 
\begin{equation}
Q_{n,1} \sim \int_{t_n}^{t_{n+1}}{D_c \partial_x C(x=x{_n}{^+},t)}dx
        \sim D_c \sqrt{t_{n+1}-t_n} 
\end{equation}
Invoking now the time and spacing laws (supposing that n is large enough), leads to
\begin{equation}
Q_{n,1} \sim D_c \sqrt{x_{n+1}{^2}-x{_n}{^2}} \sim D_c \sqrt{p^2+2p}x_n \approx D_c \sqrt{2p}x_n 
\end{equation}

We go on then by examining the contribution $Q_{n,2}$ associated to the flux into band $n$ at times $t>t_{n+1}$.
It turns out that once the next band appears at $x=x_{n+1}$, where the $C$ profile has its maximum at $t=t_{n+1}$, the left side of the front is trapped between two absorbing boundaries. At $t=t{_n}^{+}$, the total quantity of trapped $C$ can be geometrically evaluated as (see figure 5):
\begin{equation}
\delta C \approx \frac{1}{2} c^*(x_{n+1}-x_n)
\end{equation}
a finite fraction of which, say $\eta$ $(\eta<1)$, will diffuse into band $n$. Hence, using once again the spacing law, we can evaluate $Q_{n,2}$ as :
\begin{equation}
Q_{n,2} = \eta \delta C \approx \frac{1}{2} \eta p c^* x_n
\end{equation}
Finally, summing up the two contributions, we obtain :
\begin{equation}
Q_n = Q_{n,1}+Q_{n,2} \sim x_n
\end{equation}
Note that typically, $Q_{n,2}$ is smaller than $Q_{n,1}$ by four order of magnitude.

The final step of our discussion deals with the relation between $Q_n$ and $w_n$. 

The basic model described by eq. (\ref{modelCbegin})-(\ref{modelCend}) produces only infinitely thin bands. If one wants to generate bands with nonzero widths, it is necessary to supplement these equations with new terms accounting for growth mechanisms. Whatever the nature and details of these mechanisms may be, one can make the following reasonable assumption : 

The terms one has to add to equations (\ref{modelCbegin}-\ref{modelCend}) in order to model growth lead, for $t \gg t_{n+1}$, to a stationary profile of the precipitate concentration $D$ inside the $n^{th}$ bands which adopts a natural scaling form:
\begin{equation}
D = A_n D ( \frac{x-x_n}{w_n} ), \quad x_n  < x < x_n+w_n
\label{Dscaling}
\end{equation} 
where $A_n$ is an amplitude which, in general, could depend on $n$.

By equating the quantity $Q_n$ (the total amount of material that has nourished the band until its completion) with the total mass of precipitate inside the same band once formed, we have :
\begin{equation}
A_n \int_{x_n}^{x_n+w_n}{dx D(\frac{x-x_n}{w_n})} = \gamma A_n w_n = Q_n \sim x_n
\end{equation} 
where gamma is a finite constant defined as
\begin{equation}
\gamma = \int_{-1/2}^{1/2}{D(z)dz}
\end{equation}
To reproduce a relation of the form $w_n \sim x_n^{\alpha}$, the amplitude $A_n$
should be of the form $A_n \sim x_n^{\beta}$.
We obtain hereby 
\begin{equation}
\alpha + \beta =1
\label{sum}
\end{equation}
Thus the experimental value $\alpha=1$ corresponds (up to logarithmic corrections) to a constant amplitude $A$. This prediction could be tested experimentally by investigating quantitatevely the internal structure of the bands.

We can now return to the two theoretical predictions existing in the literature~\cite{luthi,dee} for the exponent $\alpha$.
In the first one, by Chopard et al., Lattice-Boltzmann cellular automata simulations including fluctuations were performed, in which microscopical mechanisms accounting for aggregation were also included, in a phenomenological way. They obtained bands patterns, which exhibited a width exponent varying between $\alpha = 0.5$ and $\alpha = 0.6$. Such a value, quite far from 1, can indeed be explained by the simple fact that their model allowed an unbounded quantity of precipitate to form at one lattice site. In this model, the amplitude $A_n$ increased  with $n$ and according to ~(\ref{sum}), constrained the exponent $\alpha$ to be smaller than 1.

In the second paper,  Dee solved a system of mean-field equations similar to (\ref{modelCbegin})-(\ref{modelCend}) in which the depletion term for the $C$ front was build from physical considerations related to the nucleation and droplet growth theories. In this model, the distribution of precipitate inside a band as obtained from numerical computation seems to follow a scaling form with an amplitude approaching a constant after some transient regime. The width exponent that could be extracted from the set of (only) six bands provided by the simulation is clearly compatible with the value $\alpha=1$.

\subsection{Summary and conclusion.}

We have reported pattern-forming experiments of Liesegang type, in which a careful measurement of the width of the bands leads to an exponent $\alpha$ close to 1. By considering the most basic theoretical model of the type related to the experiments we performed, we could explain, through a simple argumentation based on conservation laws and macroscopic properties of the concentration profiles, how such a value close to unity can be understood, and which condition it implies on the shape of the precipitate profile inside bands. 
However, only a first principle derivation, based on the underlying microscopical mechanisms of coarsening could give a clear cut determination of
$\alpha$.

We believe therefore that the microscopical theory of band formation and growth in Liesegang experiments constitutes an extremely rich and still unexplored subject of investigation in itself, for which lots remains to be done.

\section*{Acknowledgments}
We thank B. Chopard and Z. R\'acz for useful discussions.
This work has been partially supported by the 
Swiss National Science Foundation 
in the framework of the Cooperation in Science and Research with CEEC/NIS,
and by the Hungarian Academy of Sciences (Grant OTKA T 019451).

\end{multicols}

\end{document}